\documentclass[%
aps,
amsmath,amssymb,
reprint,%
]{revtex4-2}

\usepackage{graphicx}% Include figure files
\usepackage{dcolumn}% Align table columns on decimal point
\usepackage{bm}% bold math
\usepackage[utf8]{inputenc}
\usepackage[T1]{fontenc}
\usepackage{mathptmx}
\usepackage{etoolbox}

\makeatletter
\def\@email#1#2{%
	\endgroup
	\patchcmd{\titleblock@produce}
	{\frontmatter@RRAPformat}
	{\frontmatter@RRAPformat{\produce@RRAP{*#1\href{mailto:#2}{#2}}}\frontmatter@RRAPformat}
	{}{}
}%
\makeatother
\begin{document}

%\preprint{AIP/123-QED}

\title[]{Terahertz radiation generation by laser-resonant excitation of 
terahertz surface magnetoplasmons on a graphene-n-InSb semiconductor interface}
\vspace{0.5cm}

\author{Rohit Kumar Srivastav$^{1}$}

%\affiliation{Institute for Plasma Research, Bhat, Gandhinagar 382428, India }%\email [Corresponding author email:]{physicsrohit93@gmail.com}
%\altaffiliation[Also at ]{Department of Physics and Materials Science and Engineering, Jaypee Institute of Information Technology, Noida, Uttar Pradesh, 201309, India}%Lines break automatically or can be forced with \\
\author{Mrityunjay Kundu$^{1,2,*}$}
\email [Corresponding author email:]{mkundu@ipr.res.in}
%\homepage{http://www.Second.institution.edu/~Charlie.Author.}
\affiliation{$^{1}$Institute for Plasma Research, Bhat, Gandhinagar 382428, India }
\affiliation{$^{2}$
	Homi Bhabha National Institute, Anushaktinagar, Mumbai - 400094, India%\\This line break forced% with \\
}%
\date{\today}% It is always \today, today,
             %  but any date may be explicitly specified

	\begin{abstract}
		We propose a method for the laser-excitation of terahertz
surface
		magnetoplasmons via the linear mode conversion of terahertz radiation on a 
		graphene sheet deposited on an n-type semiconductor in presence of an
		external magnetic field parallel to the semiconductor surface. An obliquely incident p-polarized laser beam interacting
		with the graphene n-InSb semiconductor surface, imparts linear oscillatory velocity to the free
		electrons. This oscillatory velocity couples with the modulated electron density 
		to generate a linear current density, which resonantly excites terahertz 
		surface magnetoplasmons. It is shown that the amplitude of terahertz surface magnetoplasmons wave
		can be tuned by adjusting the external magnetic field ($\text{B}_{0}$), the 
		graphene's Fermi energy ($\text{E}_\text{F}$), the semiconductor's temperature 
		(T), and the incident angle ($\theta$) of laser. This mechanism has the potential to
		enable the development of an actively tunable plasmonic device.  %
	\end{abstract}
	
% 	\keywords{Terahertz, Graphene, n-InSb, Temperature, Surface 
% magnetoplasmon}%Use showkeys class option if keyword
	%display desired
	\maketitle
	
	\section{Introduction}\label{sec1}
 \vspace{-0.25cm}	
	Terahertz (THz) radiation, a kind of electromagnetic (EM) radiation, spans 
	frequencies between $0.1-10$~THz. Thus it is placed between the
	infrared and 
	microwave regions of the EM spectrum with wavelength~\cite{tonouchi2007cutting,jepsen2011terahertz} from 1 mm to 30 $\mu
	m$. A unique feature of THz 
	radiation is its ability to penetrate materials like plastic, paper, clothing, 
	and wood materials that are typically opaque to visible lights with minimal 
	energy loss. Many molecules are detectable at THz frequencies due to the strong 
	rotational and vibrational resonances they exhibit~\cite{zhang2010introduction}. 
	Terahertz technology is now applied in many areas, e.g. explosive and drug
	detection~\cite{wekalao2024graphene,bandurin2018resonant}, security 
	screening~\cite{tzydynzhapov2020new,liu2007terahertz}, 
	spectroscopy~\cite{zhang2023terahertz,koch2023terahertz} and medical 	research~\cite{wekalao2024terahertz,gezimati2023advances,gezimati2023terahertz,
		gong2020biomedical}. Unlike x-rays, which pose health risks, THz devices are 
	known to be
	safe for non-destructive testing~\cite{zhang2010introduction}. 
	
	Radiation of THz frequencies can be generated through linear and 
	nonlinear processes, such as optical 
	rectification~\cite{Shukla2024,Hedegaard2023,guiramand2022near}, the beating of 
	two lasers~\cite{singh2024laser,zhang2024generation,ghayemmoniri2023terahertz}, 
	and harmonic generation~\cite{di2024compact}, among others. It can also be 
	produced by lasers (or electron beams) interacting with plasmas, 
	metals, semiconductors, or carbon 
	nanotubes~\cite{VARSHNEY2022168353,choobini2023variable,sun2022terahertz,
		slepyan2001high,PhysRevLett.98.026803}. Hua \emph{et 
		al.}~\cite{PhysRevAccelBeams.27.081301} investigated THz generation by
	intense laser interaction with two solid layers using
	particle-in-cell simulation. Chamoli \emph{et al.}~\cite{chamoli2024terahertz}
	studied THz generation over metallic surface by beating of two
	laser beams in the presence of an external magnetic field. In this case, 
	reported THz frequency ranges $\sim1-4.5$~THz with magnetic fields $10-30$~T. Also Vij~\cite{Vij2024JAP} proposed THz generation
	by Gaussian laser
	beam interaction with a magnetized carbon nanotube, where magnetic fields of
	$200-500$~T were used. Kumar \emph{et
		al.}~\cite{kumar2016linear} explored the effect of external magnetic field over 
	linear mode conversion (LMC) of THz radiation into THz surface plasma waves 
	(SPWs) on a rippled n-InSb semiconductor surface. Recently, Srivastav and 
	Panwar~\cite{SrivastavPanwar+2023+572+578} reported THz generation 
	over a graphene (a two-dimensional material with carbon
	atoms organized in a hexagonal lattice) surface by using LMC of THz radiation into THz SPWs.
	
	The importance of graphene is recognized due to its distinct
	thermal, optical, mechanical, and electrical characteristics; and there has 
	been a significant increase in scientific interest in 
	it~\cite{novoselov2005two,TAO2018249,gonccalves2016introduction}. Graphene 
	holds promise for numerous applications in two-dimensional photonics and 
	plasmonics~\cite{huang2017graphene,chen2017flatland,mikhailov2007new}. It can 
	sustain localized EM surface plasmon polariton waves with both TE and TM 
	polarization~\cite{low2017polaritons,kuzmin2016transverse,teng2020graphene, 
		moiseenko2023terahertz}. Graphene-based structures enable strong light-matter 
	interactions owing to their high confinement and extended propagation 
	length~\cite{koppens2011graphene,MinovKoppensich:11}. They facilitate the 
	excitation and transmission of surface plasmons (SPs) within the THz spectrum. 
	In contrast to traditional plasmonic materials such as noble metals, graphene 
	exhibits plasmonic resonance within the crucial THz frequency 
	range~\cite{kim2018amplitude,low2014graphene}. Compared to precious metals, 
	graphene SPs demonstrate much stronger confined light field. Furthermore, the 
	properties of graphene SPs can be tailored by varying the electrostatic gate 
	voltage or applying chemical doping~\cite{ye2017broadband,yao2018broadband}. 
	Thus, graphene SPs have been employed in a variety of applications, such as 
	sensors~\cite{wu2018plasmon}, optical modulators~\cite{hao2019experimental}, and 
	devices for controlling polarization~\cite{hu2017ultrabroadband}. 
	Metals such as gold and silicon are crucial for SPs in the visible range, but 
	their role is less significant in the THz range due to their high plasma 
	frequencies. A suitable {\em alternative} for exploring SPs in the THz frequency range
	is the n-InSb semiconductor and graphene (as in this work), which features an electron density on
	the order of $10^{23} m^{-3}$ and $10^{12} m^{-2}$ respectively, with a 
	corresponding plasma frequency $\omega_{p}$ within the THz range. The strong 
	influence of external magnetic fields on its dispersion relation and the ability 
	to adjust electron density with temperature enhance the potential of 
	semiconductors for generating THz SPs~\cite{Srivastav_Panwar_2024}. Notably, the 
	generation of THz radiation over a graphene-n-InSb semiconductor rippled surface under 
	an external magnetic field using LMC has not yet been investigated in detail.
	
	In the present article, we explore the effects of an external magnetic 
	field, the temperature of the n-InSb semiconductor, and the Fermi energy of
	graphene on the excitation of THz surface magnetoplasmons (SMPs) wave through 
	the interaction of a p-polarized incident laser beam with a rippled 
	graphene-n-InSb 
	semiconductor surface. The external magnetic field is assumed parallel to the semiconductor surface. The graphene layer on the
	n-InSb semiconductor rippled surface may be developed by chemical vapour 
	deposition or plasma enhanced chemical
	vapour deposition method~\cite{liu2022achievements,saeed2020chemical}. 
	Here, graphene provides additional 
	conductivity $\sigma_g$ that supports strong SPWs, and the rippled surface 
	provides the extra wave number enabling resonant 
	excitation of THz SMPs wave in a magnetic field. The present 
configuration with the surface-parallel applied external magnetic field is similar to 
that considered in the seminal works~\cite{liu2015directional,YuPRL2008}, 
exhibiting asymmetric wave dispersion. We take the advantage of this effect for 
the  excitation of THz SMPs. However, contrary to the earlier 
works~\cite{liu2015directional,YuPRL2008} reported in the relatively high 
magnetic field and high Fermi-energy regime, we find a parameter space spanning 
the low magnetic field and low Fermi-energy where respective dispersion curves 
move in the reverse direction with increasing the magnetic field and Fermi 
energy, compared to those in Refs.~\cite{liu2015directional,YuPRL2008} w.r.t. 
the reference light-line dispersion. This leads to generation of low frequency 
THz radiations below $\sim 6$~THz which are useful in many applications. 
Additionally, our main findings
			indicate that the amplitude of THz SMPs is tunable by the Fermi energy of
			graphene, the external magnetic field, the temperature of n-InSb, and the
			incidence angle of the laser.
	
	The layout of this 
	article is as follows: we derive the expression of linear current density in 
	Sec.\ref{sec2} and
	the wave amplitude of THz SMPs in Sec.\ref{sec3}. Discussion on results and
	conclusion are given in Sec.\ref{sec4} and Sec.\ref{sec5} respectively.
	\begin{figure}
		\centering
		\includegraphics[width=8cm]{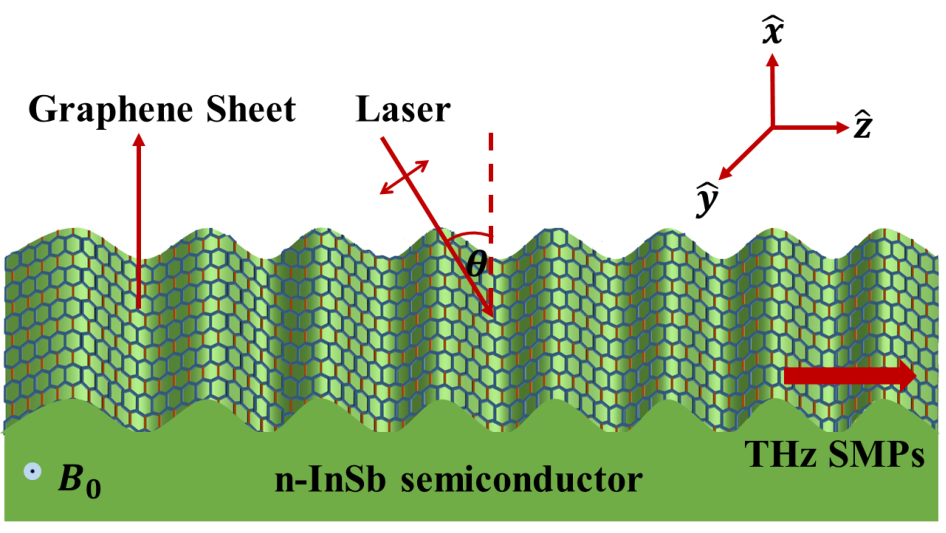}% Here is how to import EPS art
\vspace{-0.25cm}		
		\caption{\label{Fig1} Schematic of an obliquely incident p-polarized laser beam interacting with a graphene sheet deposited on n-type semiconductor surface in the presence of an external magnetic field $\text{B}_{0} \hat{y}$. The incident angle is $\theta$ w.r.t. to the surface normal (dashed line).}
			\label{Fig1}
\vspace{-0.25cm}			
	\end{figure}
	
\vspace{-0.25cm}	
	\section{Linear current density: the source of TH\lowercase{z}}\label{sec2}
\vspace{-0.25cm}	
	We assume that 
	the graphene-n-InSb semiconductor rippled surface is modulated with the 
	perturbed electron density $n=(n_{0}/2)+n_{q}$, $qa\ge1$ where 
	$n_{q}=(n_{0}/2)\cos qz$, $q$ is ripple wave number and $a$ is ripple 
	amplitude~\cite{kumar2016linear}. The graphene
	conductivity is $\sigma_g = (i {\mathrm{e}}^2 \text{E}_\text{F}) / (\pi \hbar^2 
	(\omega_{0} + i \nu))$ in the THz
	frequency regime where graphene's Fermi energy $\text{E}_\text{F} \gg h 
	\omega_{0}$~\cite{garcia2014graphene,xu2021optical}, $\mathrm{e}$ is the
	charge of an electron, $\hbar$ is 
	the reduced Planck constant, $\omega_{0}$
	is the incidence light frequency and $\nu$ is the average collision frequency 
	of electrons. In an
	external magnetic field $B_0 \hat{y}$, the permittivity of n-InSb semiconductor 
	becomes a
	tensor $\overset{=}{\epsilon}$~\cite{brion1972theory}. It components are
	$\epsilon_{x x} = \epsilon_{z z} = \epsilon_r - \epsilon_r (\omega_p^2 (\omega_{0}
	+ i \nu)) / \omega_{0} ((\omega_{0} + i \nu)^2 - \omega_{{ce}}^2)$, $\epsilon_{x
		z} = - \epsilon_{z x} = - i \epsilon_r (\omega_p^2 \omega_{{ce}}) /
	\omega_{0} ((\omega_{0} + i \nu)^2 - \omega_{{ce}}^2)$, $\epsilon_{y y} =
	\epsilon_r - \epsilon_r (\omega_p^2 / \omega_{0} (\omega_{0} + i
	\nu))$, where $\omega_{{ce}} = e B_0 / m^{\ast}_e$ is the electron cyclotron
	frequency, $\omega_p = \sqrt{n_0 e^2 / m^{\ast}_e \epsilon_r
		\epsilon_0}$ is the electron plasma frequency, $m^{\ast}_e = 0.014 m_e $ 
	is the
	effective mass of a free electron, $\epsilon_r$ is the permittivity of n-InSb
	semiconductor, $n_0 = 5.76 \times 10^{20} T^{3 / 2} \exp(0.26/k_{B} T)$ is the free electron 
	charge density, $k_B$ is the
	Boltzmann constant and $T$ is the temperature of n-InSb semiconductor~\cite{jing2022thermally, gao2023multifunctional}. Here, $\omega_{0}>\omega_{p}$ so plasma is under-dense.
	
	A p-polarized laser of frequency $\omega_{0}$ and wave numbers 
	$k_{0x}$,
	$k_{0z}$ along $\hat{x}$ and $\hat{z}$ respectively, impinges at
	an angle $\theta$ onto the rippled surface of graphene-n-InSb semiconductor  as 
	illustrated in Figure~\ref{Fig1}. The
	external magnetic field of strength $B_0$ acts along $\hat{y}$ over the
	n-InSb semiconductor. The corresponding electric field of 
	amplitude $E_0$ obeys
	\begin{equation}
		\vec{E} = E (\hat{z} + \tan \theta \hat{x}) e^{- i (\omega_{0} t + k_{0 x} x -
			k_{0 z} z)}\label{eq1} 
	\end{equation}
	
	\noindent
	where $E = E_0 \cos \theta$, $k_{0 x} = k_0 \cos \theta$, 
	$k_{0 z} = k_0 \sin
	\theta$, $k_0 = \omega_{0}/ c = \sqrt{k_{0x}^{2} + k_{0z}^{2}}$ is the 
	free space amplitude of wave vector, and $c$ is 
	the light speed.
	%
	% Upon incidence of a p-polarized laser on the graphene-n-InSb surface, the wave 
	%
	The incident light is partially transmitted and reflected. Under these 
	conditions, the resultant electric field of the transmitted wave can be 
	expressed as 
	\begin{equation}
		\vec{E}_T = E_{0} T_{tr} (\hat{z} + \beta \hat{x}) e^{\alpha x} e^{- i (\omega_{0} t -
			k_{0 z} z)}\label{eq2}
	\end{equation}
	
	\noindent 
	where $\beta = (\epsilon_{x z} \alpha  + \epsilon_{x x} i k_{0 z}) / 
	(-
	\epsilon_{x x} \alpha + \epsilon_{x z} {i k_{0 z}} )$ for $x \leqslant 0$,
	$\alpha^2 = k^2_{0 z} - (\omega^2 / c^2) \epsilon_{eff}$,
	$\epsilon_{eff} = (\epsilon^2_{x x} + \epsilon_{x z}^2) /
	\epsilon_{x x}$ and $T_{tr} = (2 \epsilon_0 \cos \theta) / ((1 +
	(\epsilon_{eff} \beta) / \tan \theta) \epsilon_0 + (\sigma_{g} \cos
	\theta) / c)$ is the transmission coefficient. Clearly, the graphene's 
	contribution enters through $\sigma_{g}$ via $T_{tr}$.
	
	The incident laser beam when interacts with the graphene-n-InSb semiconductor
	surface, ionizes their atoms, and creates free electrons. Each of these 
	electrons acquires a linear
	oscillatory velocity (in the first order approximation)
	\begin{equation}
		\vec{V} = \frac{e}{m_e^{\ast}} (\tilde{\text{v}}_{x}  \hat{x} +
		\tilde{\text{v}}_{z}  \hat{z}) {T_{tr} E_0}  e^{\alpha x} e^{- i \left( \omega_{0}
			{t - k_{0 z}}  z \right)}\label{eq3}
	\end{equation}
	where, $\tilde{\text{v}}_{x} = (\omega_{ce} - i (\omega_{0} + i \nu) 
	\beta) / ((\omega_{0} + i \nu)^2 - \omega_{ce}^2)$ and
	$\tilde{\text{v}}_{z} = (\omega_{ce} \beta - i (\omega_{0} + i \nu) ) /
	((\omega_{0} + i \nu)^2 - \omega_{ce}^2)$.
	The velocity $\vec{V}$ couples with the modulated electron
	density $n_q$, leading to an oscillatory electron current
	density %$\vec{J}_{\omega}^{l} = - (1 / 2) n_q e \vec{V}$
	\begin{equation} \vec{J}_\omega^{l} = - (1 / 2) n_q e 
		\vec{V} = (\tilde{J}_{\omega}^x
		\hat{x} + \tilde{J}_{\omega}^z \hat{z}) {T_{tr} E_0}  e^{\alpha x} e^{- i \left(
			\omega  {t - k_z}  z \right)}
	\end{equation}
	that propagates with a modified 
	propagation constant
	$k_z {= k_{0 z}}  + q $, and frequency $\omega  {= \omega 
	}_0$. The components of $\vec{J}_\omega^{l}$ 
	are $\tilde{J}_{\omega}^x=-(1/2)((n_0 
	e^2)/m_e^{\ast})\tilde{\text{v}}_{x}$, $\tilde{J}_{\omega}^z=-(1/2)((n_0 
	e^2)/m_e^{\ast})\tilde{\text{v}}_{z}$. This current density $\vec{J}_{\omega}^{l}$ acts as a
	source of THz.
	
 \vspace{-0.5cm}	
	\section{Terahertz graphene surface magnetoplasmons wave: dispersion, 
		amplitude}\label{sec3}
 \vspace{-0.25cm}	
	The electron current density $\vec{J}_\omega^l$ evolves THz SMPs at a frequency
	$\omega  {= \omega }_0$ and propagation constant $k_z {= k_{0 z}}  + q $ in the
	rippled region. The current source localized in the rippled region may be 
	represented by a delta
	function $\delta(x)$~\cite{10.1063/1.2795575}. Using the Faraday's 
law 
	$\vec{\nabla} \times
	\vec{E} = - (\partial \vec{B} / \partial t)$ and Ampere's law
	$\vec{\nabla} \times \vec{B} = \mu_0 \vec{J}_{\omega}^{l} + \mu_0 \epsilon_0 \epsilon_r
	(\partial \vec{E} / \partial t)$, the equation governing the THz SMPs wave can 
	be given as
	\begin{equation}
		\vec{\nabla}^2  \vec{E} - \vec{\nabla} (\vec{\nabla} \cdot \vec{E}) -
		\frac{\omega^2}{c^2} \left( \overset{=}{\epsilon} \vec{E}_{\omega} \right) =
		- \mu_0 i \omega \vec{J}_{\omega}^l h \delta (x).\label{eq5}
	\end{equation}
	Here, $h$ is the ripple height, $\overset{=}{\epsilon}$ is the effective 
	permittivity of n-InSb semiconductor
	(at a frequency $\omega$) with components $\epsilon_{xx} = \epsilon_{zz} = 
	\epsilon_r - \epsilon_r \omega_p^2
	(\omega + i \nu) / [\omega ((\omega + i \nu)^2 - \omega_{ce}^2)]$,
	$\epsilon_{xz} = - \epsilon_{zx} = - i \epsilon_r \omega_p^2 \omega_{ce} /
	[\omega ((\omega + i \nu)^2 - \omega_{ce}^2)]$, $\epsilon_{yy} = \epsilon_r -
	\epsilon_r \omega_p^2 / [\omega (\omega + i \nu)]$, $\epsilon_{xy} =
	\epsilon_{yx} = \epsilon_{yz} = \epsilon_{zy} = 0$ and $\epsilon_{eff}=(\epsilon^2_{x z} + \epsilon^2_{x x})/\epsilon_{x x}$.
	From Eq.~\eqref{eq5} we obtain
	\begin{multline}
		\frac{\partial^2 E^z_{\omega}}{\partial x^2} - \left( k_z^2
		{- \frac{\omega^2}{c^2} \epsilon_{eff}} \right) E^z_{\omega} = - \frac{c^2 \mu_0 i}{\omega \epsilon_{x x}
		} \Bigg[ \left({\frac{\omega^2}{c^2} \epsilon_{x x}} - k_z^2
		\right) \tilde{J}_{\omega}^z \\ + \left(
		{\frac{\omega^2}{c^2} \epsilon_{x z}} + i k_z \frac{\partial}{\partial_x}
		\right) \tilde{J}_{\omega}^x \Bigg] h \delta (x)
		\label{eq6}
	\end{multline}
	Making right-hand side of Eq.~\eqref{eq6} zero, the electric field for THz
	SMPs can be expressed as 
	\begin{equation}
		\vec{E} = E_l  \mathrm{e}^{- i (\omega  t - k_z z)} \left\{\begin{array}{ll}
			\left( \hat{z} {+ \beta_1}   \hat{x} \right) e^{- \alpha_1 x}, &
			\textrm{air x} > 0\\
			\left( \hat{z} {+ \beta_2}   \hat{x} \right) e^{{\alpha }_2 x}, & 
\hskip -0.25cm
			\textrm{graphene-n-InSb x} \leq 0
%			\textrm{x} \leq 0
		\end{array}\right.\label{eq7}
	\end{equation}
	where, $E_l$ is the amplitude of THz SMPs wave, $\beta_1=-ik_z/\alpha_1$, 
	$\alpha_1^2=k_z^2-(\omega^2/c^2)$, 
	$\beta_2=\left(\epsilon_{xz}\alpha_2+\epsilon_ {xx}ik_z\right)/\left(-\epsilon_{ 
		xx}\alpha_2+\epsilon_{xz}ik_z\right)$, and 
	$\alpha_2^2=k_z^2-(\omega^2/c^2)((\epsilon^2_{xz}+\epsilon^2_{xx})/\epsilon_{xx} 
	)$. Here $\alpha_{1}$ and $\alpha_{2}$ are decaying constants.
	From $\vec{\nabla} \times \vec{E} = - (\partial
	\vec{B} / \partial t)$, the associated magnetic field of THz SMPs can be 
	written as
	\begin{equation}
		{\vec{H} = E_l  e^{- i (\omega  t - k_z z)} \hat{y}
			\left\{\begin{array}{ll}
				i \frac{\omega \epsilon_0}{\alpha_1 } e^{- \alpha_1 x}, & \textrm{air
					x} > 0\\
				\frac{i \epsilon_0 \omega \epsilon_{eff} \epsilon_{x x}}{-
					\epsilon_{x x} {\alpha_2}  + \epsilon_{x z} {i k_z} } e^{{\alpha }_2 x}, &
				\textrm{graphene-n-InSb x} \leq 0\label{eq8}
			\end{array}\right.}
	\end{equation}
	Using the jump conditions $\vec{H}_{2, y} - \vec{H}_{1, y} =
	\vec{J}_{\sigma_g}$ at $x = 0$, where $\vec{J}_{\sigma_g} = \sigma_g 
	\vec{E}_z$; the dispersion relation of THz SMPs reads
	as~\cite{liu2015directional}
	\begin{equation}
		\frac{\epsilon_{eff} \epsilon_{x x}}{\epsilon_{x x} {\alpha_2}  -
			\epsilon_{x z} {i k_z} } + \frac{1}{\alpha_1} = \frac{\sigma_g}{i \omega
			\epsilon_0}\label{eq9}
	\end{equation}
	For convenience, from Eq.\eqref{eq9}, it is customary to plot the wave
dispersion curves with
	the normalized THz frequency $\Omega(=\omega/\omega_{p})$ vs normalized wave number $K_{z}(=k_{z}c/\omega_{p})$.
	\begin{figure}
		\centering
		\includegraphics[width=0.45\textwidth,
height=0.265\textheight]{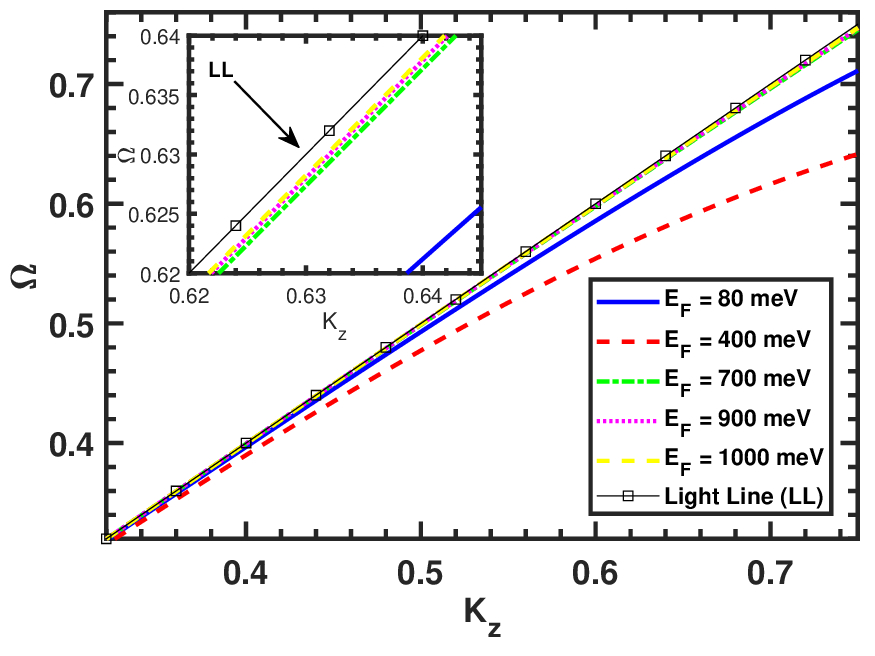}
%\cite{liu2015directional}
\vspace{-0.35cm}
		\caption{
		Wave dispersion curves with the normalized THz frequency $\Omega(=\omega/\omega_{p})$ vs normalized wave number $K_{z}(=k_{z}c/\omega_{p})$ for different values of graphene Fermi energy $\text{E}_\text{F}$ = $80-1000$ meV at a fixed value of normalized electron cyclotron frequency $\Omega_{ce}(= \omega_{ce} / \omega_p) = 0.1$ (747 Gauss). Line with $(\square)$ is the dispersion curve of EM wave in vacuum (light-line, LL). Inset shows an enlarged view for clarity.
\vspace{-0.5cm}
		}
		\label{Fig2}
	\end{figure}
Figure~\ref{Fig2} shows the dispersion curves of THz SMPs for different
values
	of
	graphene Fermi energy $\text{E}_\text{F}$ = $80-1000$~meV at a normalized
	value of electron cyclotron frequency $\Omega_{ce}(= \omega_{ce} / \omega_p) = 0.1$ (747 Gauss). It is found that dispersion curves {\emph {first move away}} from the dispersion curve of the EM wave in free space (called the light-line, LL) with {\emph {increasing}} $\text{E}_\text{F}$, then {\emph {reverses toward}} the LL for higher value of  $\text{E}_\text{F} \rightarrow 1000$~meV.
	Results with higher values of $\text{E}_\text{F}  \rightarrow 1000$~meV are similar to those in Refs.\cite{liu2015directional}, particularly above $\text{E}_\text{F} \sim 400$~meV.

	We now concentrate on results with lower values of graphene Fermi energy $\text{E}_\text{F}$ and magnetic fields.
	\begin{figure}
		\centering                                         
		\includegraphics[width=0.45\textwidth, 
height=0.265\textheight]{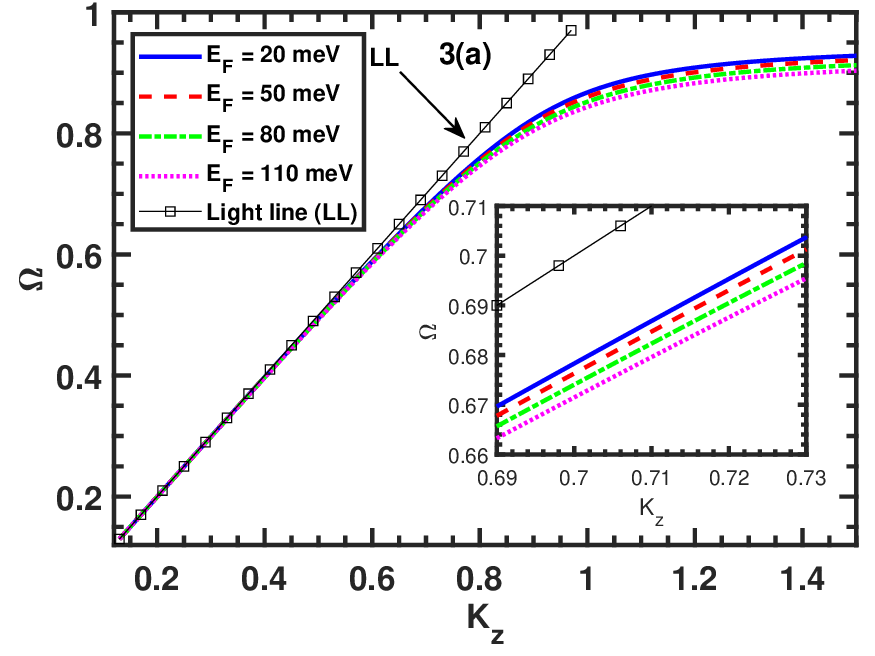}% Here is how to import EPS art

\includegraphics[width=0.45\textwidth, 
height=0.265\textheight]{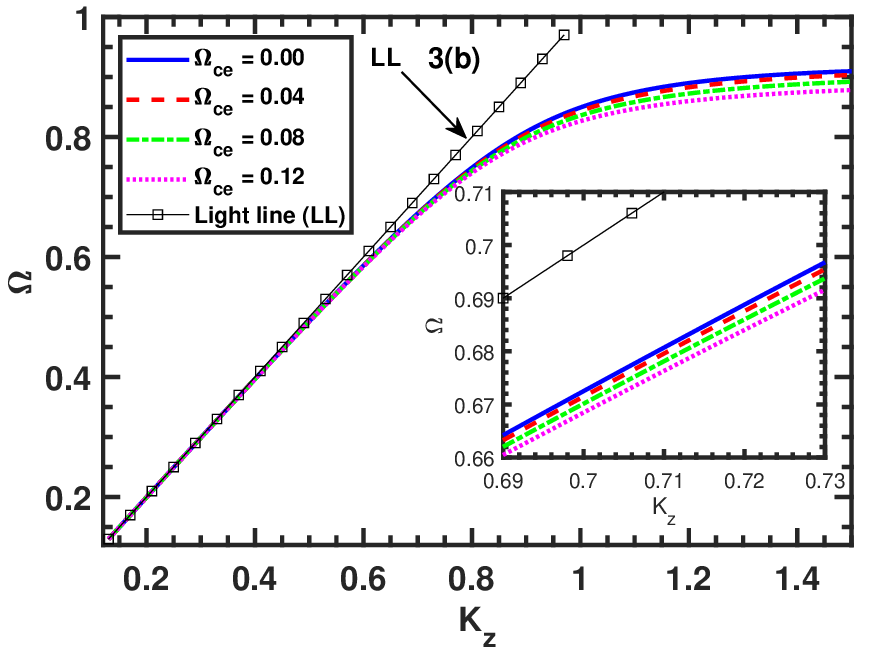}
\vspace{-0.35cm}
		\caption{
		Wave dispersion curves with the normalized THz frequency $\Omega(=\omega/\omega_{p})$ vs normalized wave number $K_{z}(=k_{z}c/\omega_{p})$: (a) for four values of graphene Fermi energy $\text{E}_\text{F} =20$ meV, $50$ meV, $80$ meV $110$ meV at a fixed value of $\Omega_{ce}(= \omega_{ce} / \omega_p) = 0.04$; and (b) for four values of $\Omega_{ce} = 0.0, 0.04, 0.08$ and 0.12  at a fixed  value of $\text{E}_\text{F} = 110$ meV. Line with $(\square)$ is the dispersion curve of EM wave in vacuum (light-line, LL). Inset shows an enlarged view for clarity.
\vspace{-0.5cm}
		}
		\label{Fig3}
	\end{figure}
	%
% 	\begin{figure}
% 		\centering                                         
% 		\includegraphics[width=0.35\textwidth, 
% height=0.2\textheight]{Figure3.eps}% Here is how to import EPS art
% 		\caption{\label{Fig3} Normalized dispersion plot of THz SMPs
% 			for four values of normalized frequency 
% 			$\Omega_{ce}$= 0.0, 0.04, 0.08 and 0.12  at a fixed  value of 
% 			graphene Fermi 
% 			energy $\text{E}_\text{F}$= $110$ meV. Black line 
% 			represents the dispersion curve of EM wave in the free space. Fig.\ref{Fig2}a shows the dispersion curves of THz SMPs for four
% different
% values
% 	of
% 	graphene Fermi energy $\text{E}_\text{F}$ = $20-110$ meV at a normalized
% 	value of electron cyclotron frequency $\Omega_{ce}(= \omega_{ce} / \omega_p) = 0.04$.
% An enlarged view of the result is shown in the inset.
% 			}
% 		\label{Fig3}
% 	\end{figure}
	%
	Figure~\ref{Fig3}a shows the dispersion curves of THz SMPs for four different values
	of $\text{E}_\text{F}$ = $20-110$ meV at a normalized
	electron cyclotron frequency $\Omega_{ce}(= \omega_{ce} / \omega_p) = 0.04$.
	Similarly, 
	Figure~\ref{Fig3}b shows the dispersion curves of THz SMPs for four
values 
	of $\Omega_{ce} = 0.0 
	-0.12$ at a fixed $\text{E}_\text{F} =
	110$ meV. It is clear that the dispersion curve of THz SMPs systematically {\emph {shifts
	away}} from the dispersion curve of the EM wave in free space (the light-line, LL) for
	{\emph {increasing}} $\text{E}_\text{F}$ and/or $\Omega_{ce}$ -- this is a feature that are opposite in nature w.r.t. those presented in Figure~\ref{Fig2} with higher values of $\text{E}_\text{F}$ and $\Omega_{ce}$.
	The
		dispersion curves (Figures~\ref{Fig3}a and \ref{Fig3}b) serve as a key analytical tool for
characterizing the
		propagation features of graphene SMPs, especially in configurations 
involving a
		n-InSb semiconductor substrate, thereby providing essential insights 
into their
		SMPs response.
	
	The excitation of TH SMPs requires phase matching condition
\vspace{-0.4cm}
	\begin{equation}
		q = k_z - k_{0 z} \label{eq10}
	\end{equation}
with the ripple wave number $q$.
% \vspace{-0.25cm}
	%
%
	\begin{figure}
		\centering                                         
		\includegraphics[width=0.45\textwidth,
height=0.225\textheight]{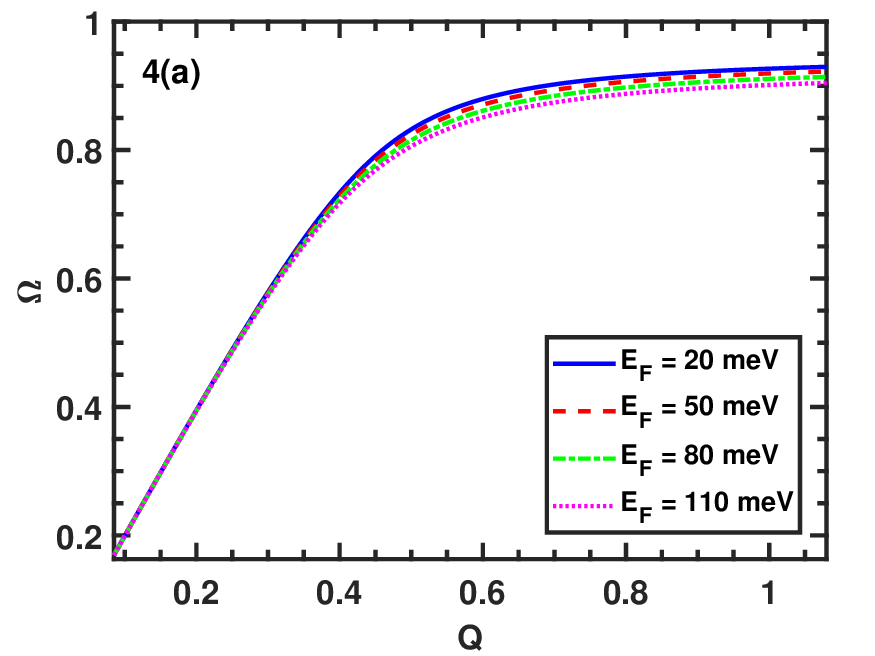}% Here is how to import EPS
		
		\includegraphics[width=0.45\textwidth,
height=0.225\textheight]{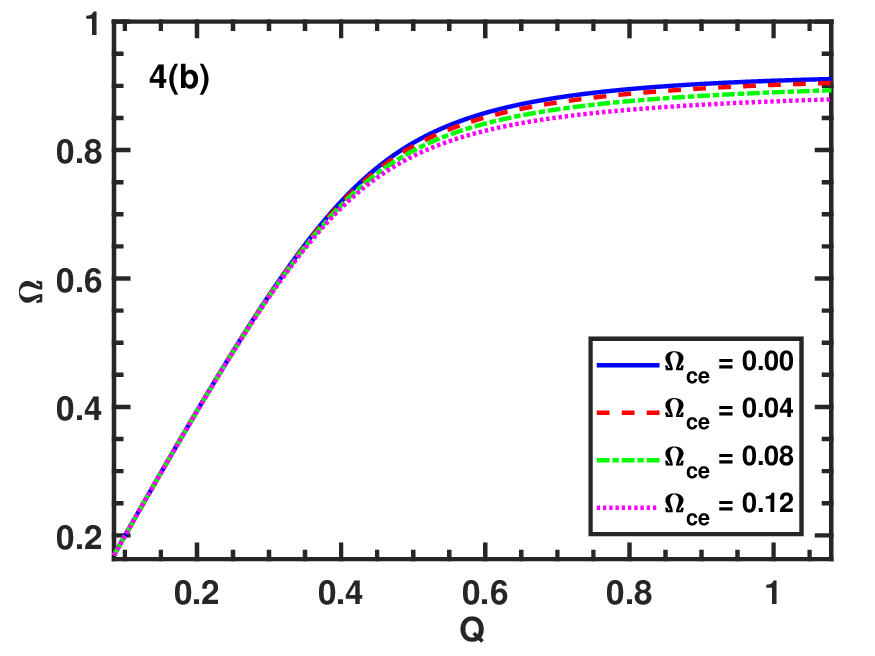}
		
		\includegraphics[width=0.45\textwidth,
height=0.225\textheight]{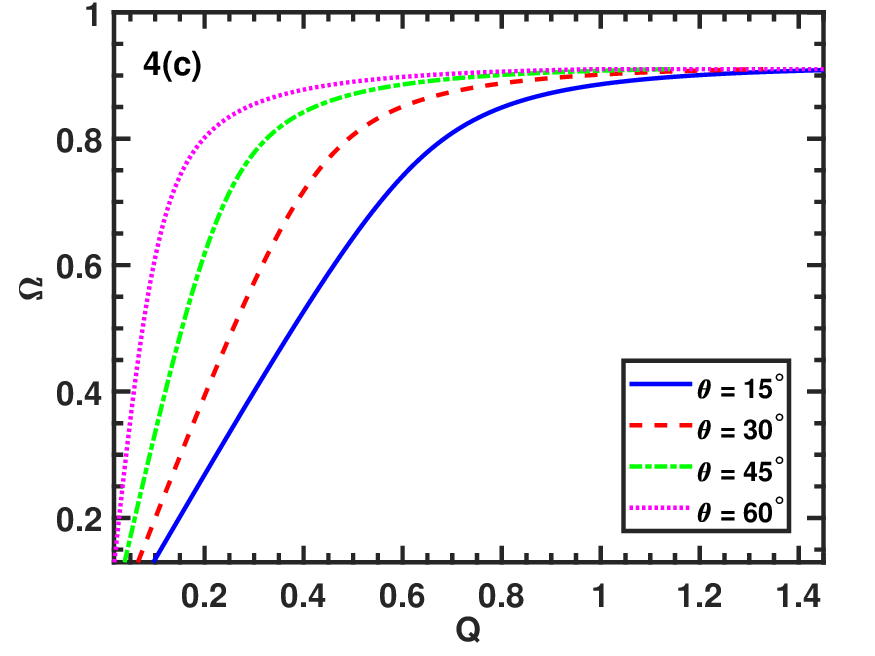}
\vspace{-0.35cm}
		\caption{Dispersion curves with normalized ripple wave number $Q=qc/\omega_{p}$ verses normalized THz SMPs frequency $\Omega (= \omega/\omega_{p})$: (a) for four values of graphene Fermi energy $\text{E}_\text{F}$ = $20$ meV, $50$ meV, $80$ meV and $110$ meV at a fixed value of incident angle $\theta = 30^{\circ}$ and $\Omega_{ce} = 0.04$; (b) for four values of $\Omega_{ce} = 0.0 -0.12 $ at a fixed value of $\text{E}_\text{F} = 110$ meV and $\theta=30^{\circ}$; and (c) for four values of $\theta = 15^{\circ}$, $30^{\circ}$, $45^{\circ}$ and $60^{\circ}$ at a fixed value $\text{E}_\text{F}$ = $110$ meV and $\Omega_{ce} = 0.04$. 
	}
		\label{Fig4}
\vspace{-0.5cm}		
	\end{figure}
	%
% 	\begin{figure}
% 		\centering                                         
% 		\includegraphics[width=5.5cm]{Figure5.eps}% Here is how to import EPS art
% 		\caption{Normalized ripple wave number
% 			$Q=qc/\omega_{p}$ verses 
% 			normalized THz SMPs frequency $\Omega (= \omega/\omega_{p})$ for 
% four  
% 			values of normalized 
% 			$\Omega_{ce}$ = 0.0, 0.04, 0.08, 0.12  at particular values of incident angle 
% 			$\theta=30^{\circ}$ and $\text{E}_\text{F}$ = $110$ meV.}
% 		\label{Fig5}
% 	\end{figure}
% 	
% 	\begin{figure}
% 		\centering                                         
% 		\includegraphics[width=5.5cm]{Figure6.eps}% Here is how to import EPS art
% 		\caption{Normalized ripple wave number
% 			$Q=qc/\omega_{p}$ verses 
% 			normalized THz SMPs frequency $\Omega (= \omega/\omega_{p})$ for 
% four 
% 			values of incident angle $\theta = 15^{\circ}$, 
% 			$30^{\circ}$, $45^{\circ}$ and $60^{\circ}$ at a fixed value of  
% 			$\text{E}_\text{F}$ = $0.110$ meV and $\Omega_{ce} = 0.04$.}
% 		\label{Fig6}
% 	\end{figure}
%
	Figure ~\ref{Fig4}a shows the normalized ripple wave number $Q=qc/\omega_{p}$
verses 
	normalized THz frequency $\Omega (= \omega/\omega_{p})$ for four different 
	values of $\text{E}_\text{F} = 20-110$ meV at a fixed 
	value of $\Omega_{ce} = 0.04$ and incident angle $\theta=30^{\circ}$.
Similarly, 
Figure~\ref{Fig4}b shows $Q$ 
	verses $\Omega$ for four different values of 
	$\Omega_{ce} = 0.0 -0.12 $
	at a fixed $\text{E}_\text{F} =
	110$ meV and $\theta=30^{\circ}$. Figure~\ref{Fig4}c represents
$Q$ verses 
	$\Omega$ for $\theta = 15^{\circ}$,
$30^{\circ}$, 
	$45^{\circ}$ and $60^{\circ}$ at a fixed $\text{E}_\text{F}$ = $110$ 
	meV and $\Omega_{ce} = 0.04$. 
	We employ the same numerical parameters that have 
been previously
		used for laser interactions with graphene or semiconductor for
		generating SPs~\cite{SrivastavPanwar+2023+572+578,kumar2016linear}. The present SMPs dispersion curves (Figures~\ref{Fig3} and \ref{Fig4}) also align with the trend observed in the dispersion curve reported by Liu \emph{et
al.}~\cite{liu2015directional} with SPs. Since $q\propto k_z$, results in Figures~\ref{Fig4}a, b are almost quantitatively similar to those in Figures~\ref{Fig3}a, b respectively. Additional feature in dispersion is brought out with the angular dependence shown in Figure~\ref{Fig4}c.
% They also reported the dispersion curve of SMPs shifts towards the dispersion curve of light line increases the magnetic field at fixed value of Fermi energy and used high magnetic field 1 T and Fermi energy 1 eV. Our results show that
% how the external magnetic
% 		field and graphene Fermi energy induce asymmetric dispersion
% curve for the SMPs.

 \vspace{-0.5cm}
	\subsection{Approximate terahertz SMP field}
 \vspace{-0.25cm}
	When the current density is finite, the right-hand side of Eq.\eqref{eq6} is
	also finite. We assume that the mode structure of THz SMPs remains constant ~\cite{kumar2016linear}, while the amplitude of THz SMPs
	varies as a function of $z$. Under these conditions, we seek for a solution of 
	the form
	\begin{equation}
		\vec{E} = E_l  (z)  \vec{\psi}  (x) e^{- i (\omega  t - k_z z)} \label{eq11}
	\end{equation}
where,
\vspace{-0.225cm}
\begin{equation}
		\vec{\psi}  (x) = \left\{\begin{array}{ll}
			\left( \hat{z} {+ \beta_1}   \hat{x} \right) e^{- \alpha_1 x}, &
			\textrm{air x} > 0\\
			\left( \hat{z} {+ \beta_2}   \hat{x} \right) e^{{\alpha }_2 x}, & 
\hskip -0.25cm
			\textrm{graphene-n-InSb x} \leq 0
%			\textrm{x} \leq 0
		\end{array}\right.\nonumber
	\end{equation}
	Letting $k_z \rightarrow (k_z - i \partial / \partial z)$, and using WKB
	approximation in Eq.\eqref{eq6} with Eq.\eqref{eq11}, we obtain
	%
%\vspace{-0.25cm}
	\begin{multline}
		2 k_z \vec{\psi}  (x) \frac{\partial  E_l }{\partial z} e^{-
			i (\omega  t - k_z z)} = - \frac{c^2 \mu_0}{\omega \epsilon_{x x} } 
		\Bigg[ \left( {\frac{\omega^2}{c^2}
			\epsilon_{x x}} - k_z^2 \right) \\ \times \tilde{J}_{\omega}^z   \hat{z} +
		\left({\frac{\omega^2}{c^2} \epsilon_{x z}} + i k_z
		\alpha_{2 \omega}  \right) \tilde{J}_{\omega}^x  \hat{x}
		\Bigg] h \delta x
		\label{eq12}
	\end{multline}
	Multiplying Eq.\eqref{eq12} by $\vec{\psi} ^{\ast} (x) d x$ and integrating
	from
	$- \infty$ to $+\infty$, we obtain
%\vspace{-0.25cm}
	%
	\begin{multline}
		2 k_z \frac{\partial  E_l }{\partial z} e^{- i (\omega  t - k_z z)} = -
		\frac{c^2 \mu_0}{\omega \epsilon_{x x} } \frac{1}{I_1} \Bigg[ I_2 \bigg(
		{\frac{\omega^2}{c^2} \epsilon_{x x}} - k_z^2
		\bigg) \\+ \bigg( {{\frac{\omega^2}{c^2}
				\epsilon_{x z}}} + i k_z \alpha_{2 \omega}  \bigg) I_3 \Bigg] 
\label{eq13}
	\end{multline}
	\noindent
	where, \ $I_1 = \int^{\infty}_{- \infty} \vec{\psi}^{\ast} (x) \cdot
	\vec{\psi}^(x) d x$, $I_2 = \int^{\infty}_{- \infty} \vec{\psi} ^{\ast}
	(x) \cdot \tilde{J}_{\omega}^z h \delta x \hat{z} d x$ and $I_3 =
	\int^{\infty}_{- \infty} \vec{\psi} ^{\ast} (x) \cdot\tilde{J}_{\omega}^{x} h \delta x 
	\hat{x} dx$.
	By integrating Eq.\eqref{eq13} concerning $z$ over the illumination length
	$d$, we determine the amplitude of the THz SMPs to be 
	
%	\begin{widetext}
		\begin{eqnarray}
		\nonumber
			\bigg| \frac{E_l }{E_0 } \bigg| & = & \bigg| \frac{ 
i \omega^2_{p}}{2 k_z \omega
				\epsilon_{x x} } {\left( \frac{ 1 + \beta^2_1 
				}{{\alpha }_1} + \frac{ 1 + \beta^2_2}{{\alpha
					}_2} \right)^{-1}} 
					\bigg[2 \left({\frac{\omega^2}{c^2}
				\epsilon_{x x}} - k_z^2 \right)\tilde{\text{v}}_{z}
			\\ & + & \left({\frac{\omega^2}{c^2} \epsilon_{x z} + i k_z
				\alpha_{2 } }\right) ( {{\beta^{\ast}_1}  {+ \beta^{\ast}_2} 
			})\tilde{\text{v}}_{x}\bigg] T_{tr} h d \bigg|. \label{eq14}
		\end{eqnarray}
%	\end{widetext}
%
	Equation~\eqref{eq14} represents the ratio of THz SMP field strength $E_l$ 
	with incident laser field strength $E_0
	$. It is directly proportional to the
	transmission coefficient $T_{tr}$, ripple height~$h$ and illumination 
length 
	$d$. Results are shown in Sec.\ref{sec4}. 
	
\vspace{-0.45cm}
	\section{Results and Discussion}\label{sec4} 
\vspace{-0.25cm}
	For illustration of variation of THz SMPs field amplitude $|E_{l}/E_{0}|$, we choose the CO$_2$ laser with wavelength $\lambda
=
	10.81 \mu m$ and intensity $I_0 = 2 \times 10^{15}\, W/\mathrm{cm}^2$. The
	collision frequency $\nu = 0$, illumination length $d = 10 \mu m$, ripple
	height $h = 10 \mu m$, n-InSb semiconductor's relative permittivity 
$\epsilon_r
	= 15.68$, electron density $n_0 = 2.4 \times 10^{23} m^{- 3}$, electron 
plasma
	frequency $\omega_p = 9.38$ THz are kept fixed. The normalized THz 
frequency $\Omega$ vary from
	0.1 to 0.5, normalized electron cyclotron frequency $\Omega_{ce}$ vary
	from 0 to 0.12, graphene Fermi energy $\text{E}_\text{F}$ vary from 20 meV 
to
	140 meV. These numerical parameters have 
also been
		used previously for laser interactions with graphene or 
semiconductor,
		relevant for generating SPs (or 
SMPs)~\cite{10.1117/1.JNP.11.036015,SrivastavPanwar+2023+572+578,
kumar2016linear} and also results in Sec.\ref{sec3}.

% 	The
% 	variation of normalized THz SMPs wave amplitude $E_l  / E_0 $ with 
% normalized
% 	THz frequency~$\Omega$ for various values of graphene's
% 	Fermi energy $\text{E}_\text{F}$, normalized electron cyclotron frequency
% 	$\Omega_{ce}$ and incident laser
% 	angle~$\theta$ are shown in Figs.\ref{Fig7}, \ref{Fig8}, \ref{Fig7} and 
% 	\ref{Fig8} respectively.	
	
\begin{figure}[]
		\centering                                         
		\includegraphics[width=0.45\textwidth,
height=0.275\textheight]{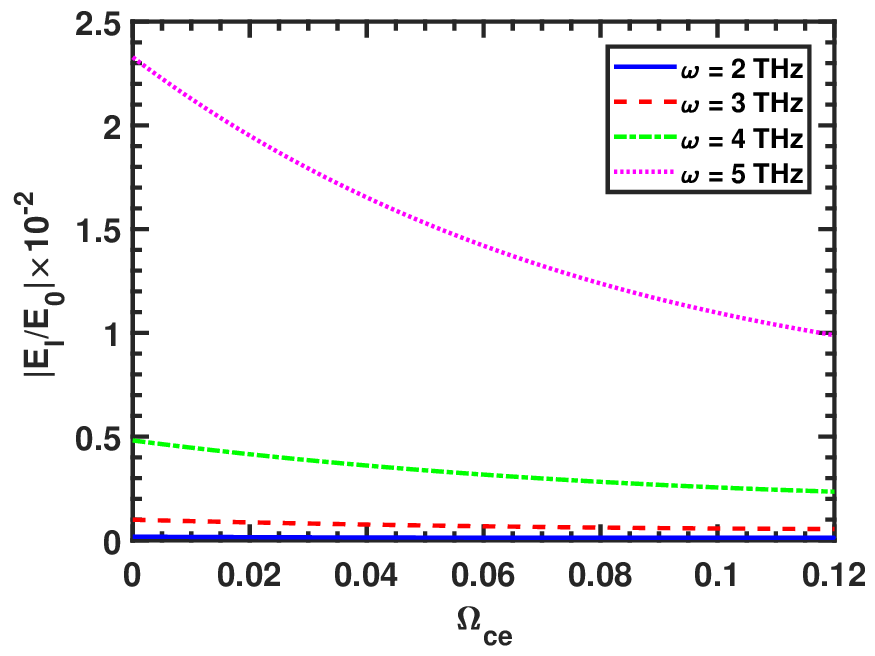}% Here is how to import EPS art
\vspace{-0.35cm}	
		\caption{Normalized THz SMPs field amplitude $|E_{l}/E_{0}|$ versus $\Omega_{ce}$ for different THz frequency $\omega$ = 2 THz, 3 THz, 4 THz and 5 THz at a fixed value of graphene Fermi energy $\text{E}_\text{F}$= $120$ meV and incident angle $\theta = 82.83^{\circ}$. Other parameters are ripple height $\text{h} = 10 \mu m$, illumination length $\text{d} = 10 \mu m$, and plasma frequency $\omega_{p} = 9.38$ THz.}
		\label{Fig5}
\vspace{-0.5cm}	
	\end{figure}

	Figure~\ref{Fig5} represents normalized THz SMPs field amplitude $|E_{l}/E_{0}|$
	verses $\Omega_{ce}$ for different values 
of THz frequency $\omega$ = 2 THz, 3 THz, 4 THz and 5 THz at 
a fixed value of $\text{E}_\text{F} 
= 120$ meV and $\theta =
	82.83^{\circ}$. It is shown that THz SMPs field amplitude 
$|E_l/E_0|$ decreases with increasing
	$\Omega_{ce}$. This is due to the restricted motion of electrons with increasing external magnetic field.
	
\begin{figure}[]
		\centering                                         
		\includegraphics[width=0.45\textwidth,
height=0.275\textheight]{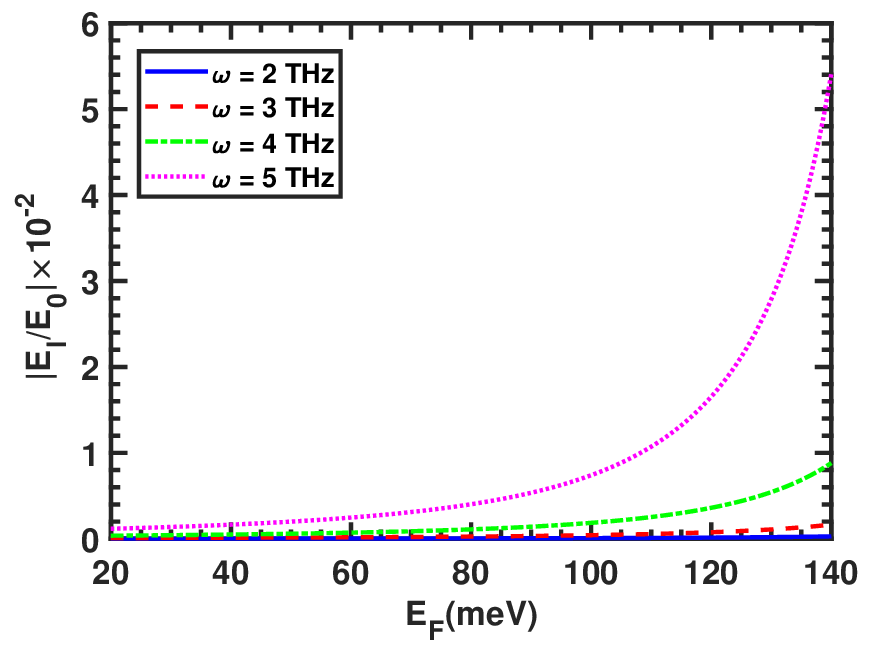}% Here is how to import EPS art
\vspace{-0.35cm}
		\caption{Normalized THz SMPs field amplitude $|E_{l}/E_{0}|$ versus graphene Fermi energy for different THz frequency $\omega$ = 2 THz, 3 THz, 4 THz and 5 THz at a fixed value of incident angle $\theta = 82.83^{\circ}$ and frequency $\Omega_{ce}= 0.04$.  
% The other parameters are ripple height $\text{h} = 10 \mu m$, illumination length $\text{d} = 10 \mu m$, plasma frequency $\omega_{p} = 9.38$~THz.
			Other parameters are same as in Fig.\ref{Fig5}.
			}
		\label{Fig6}
	\end{figure}

	Figure~\ref{Fig6} shows
 $|E_{l}/E_{0}|$ verses 
$\text{E}_\text{F}$ for
	different values of THz frequency $\omega$ = 2 THz, 3 THz, 4 THz and 5 THz
	at a fixed $\Omega_{ce}=0.04$ and
	$\theta = 82.83^{\circ}$.  The THz SMPs field amplitude
$|E_l/E_0|$ is shown to increase with
	increase in $\text{E}_\text{F}$ for each $\omega$.
	
\begin{figure}[t]
		\centering                                         
		\includegraphics[width=0.45\textwidth,
height=0.275\textheight]{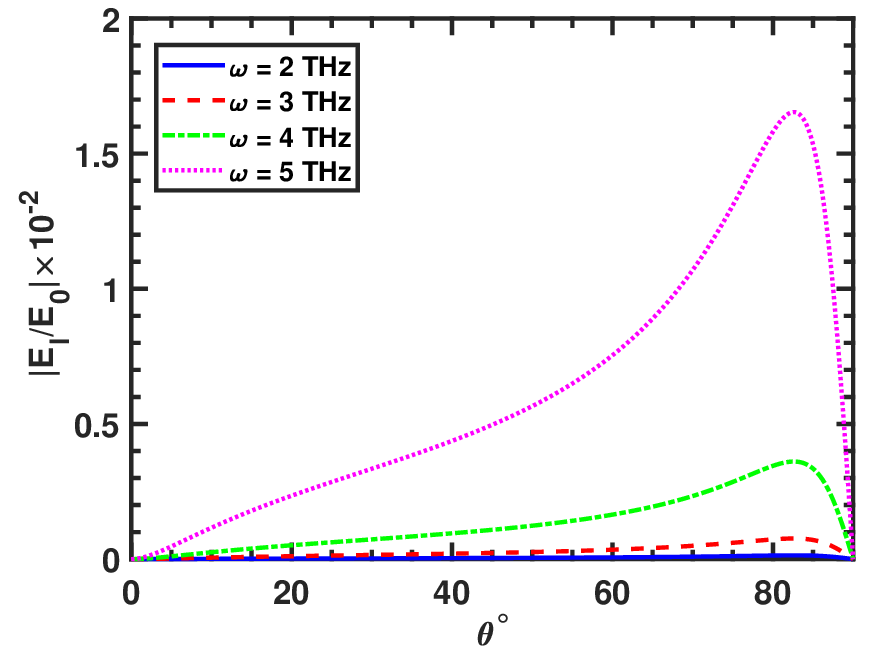}% Here is how to import EPS art
\vspace{-0.25cm}
		\caption{Normalized THz SMPs field amplitude $|E_{l}/E_{0}|$ versus incident laser angle $\theta$  for different THz frequency $\omega$ = 2 THz, 3 THz, 4 THz and 5 THz at a fixed value of $\text{E}_\text{F}$ = $120$ meV and $\Omega_{ce}= 0.04$. 
% The other parameters 
% are ripple height $\text{h} 
% = 10 \mu m$, illumination length $\text{d} = 10 \mu m$, plasma frequency 
%$\omega_{p} = 9.38$ THz.
			Other parameters are same as in Fig.\ref{Fig5}.
\vspace{-0.5cm}				}
		\label{Fig7}
	\end{figure}

	Figure~\ref{Fig7} shows
 $|E_{l}/E_{0}|$ verses
 incident angle $\theta$ for
	different values of THz frequency $\omega$ = 2 THz, 3 THz, 4 THz and 5 THz
	at a fixed $\Omega_{ce}=0.04$ and $\text{E}_\text{F} = 120$ meV. 
The THz SMPs field amplitude 
$|E_l/E_0|$ increases 
	with increasing $\theta$, reaches a maximum for $\theta\rightarrow 
90^{\circ}$ (grazing incidence) depending upon the $\omega$ values, and then it 
decreases with further increase in $\theta$. It suggests that effects are 
pronounced near $\theta\approx 
83^{\circ}$ due to linear resonance. At normal incidence no THz is generated. Thus an angle $\theta = 82.83^{\circ}$ is chosen in Figures~\ref{Fig5} and \ref{Fig6} to obtain the maximum effect.
	
\begin{figure}
		\centering                                         
		\includegraphics[width=0.45\textwidth,
height=0.275\textheight]{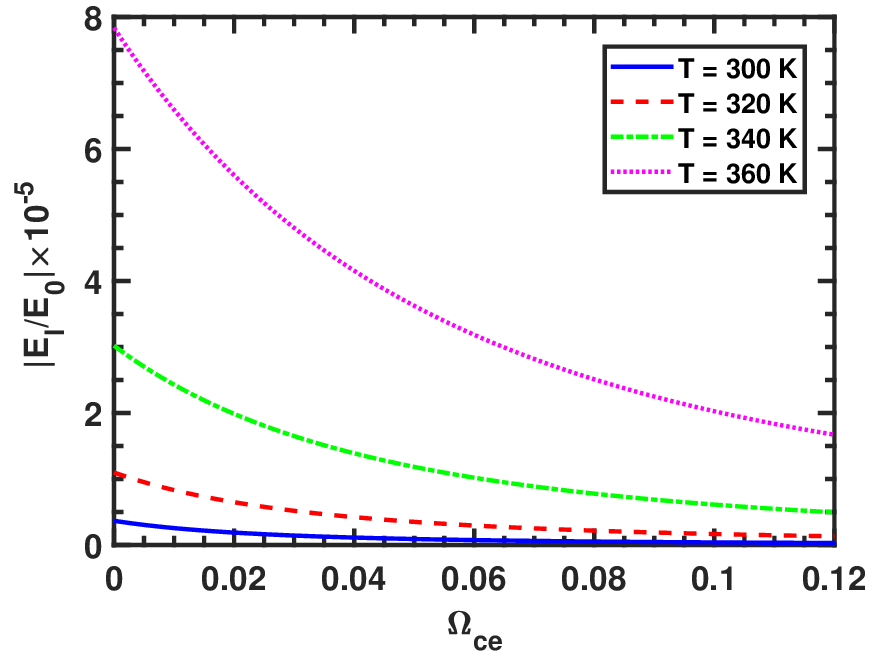}% Here is how to import EPS art
\vspace{-0.25cm}
		\caption{Normalized THz SMPs field amplitude $|E_{l}/E_{0}|$ versus frequency $\Omega_{ce}$ for different values of n-type semiconductor temperature T $=300$ K,  $320$ K,  $340$ K and  $360$ K at a fixed value of $\text{E}_\text{F}$ = $120$ meV, THz frequency $\omega = 5 $THz and incident laser angle $\theta = 82.83^{\circ}$. 
% 			The other parameters are ripple height $\text{h} = 10 \mu m$ and 
% illumination length 
% 			$\text{d} = 10 \mu m$.
Other parameters are same as in Fig.\ref{Fig5}.
\vspace{-0.35cm}				}\label{Fig8}
	\end{figure}

	Figure~\ref{Fig8} shows
 $|E_{l}/E_{0}|$ verses $\Omega_{ce}$ for four different values of
	n-InSb semiconductor temperature T=300 K, 320 K, 340 K and 360 K with a 
fixed value 
of
	$\text{E}_\text{F} = 120$ meV, THz frequency 
$\omega=5$ THz and
	 $\theta = 82.83^{\circ}$. The THz SMPs field amplitude 
$|E_{l}/E_{0}|$  
decreases with increase 
in $\Omega_{ce}$ for $T=300-360$~K as in Figure~\ref{Fig5}. It suggests that a low
magnetic field and higher temperature may lead to higher THz field intensity.
	It has been previously reported~\cite{Srivastav_Panwar_2024} that use of a p-polarized laser beam in an optical rectification mechanism results in THz SPs radiation with a normalized amplitude $|E_{l}/E_{0}| \approx 10^{-4}$. In the present work we use LMC process to generate THz radiation on a graphene-n-InSb semiconductor surface in the presence of an external magnetic field, leading to almost 100 times increase in the field amplitude (as in Figures~\ref{Fig5}, \ref{Fig6}, \ref{Fig7}). Our results are also found to be
consistent
		with earlier reported THz SPs amplitudes for the interaction of 
s-/p-polarized laser beams with graphene and rippled n-InSb 
semiconductor
		surface, 
respectively~\cite{SrivastavPanwar+2023+572+578,kumar2016linear}.

\vspace{-0.5cm}
	\section{Conclusion}\label{sec5}
\vspace{-0.25cm}
	In this work, we have explored the excitation of THz SMPs wave
by analyzing the interaction of a p-polarized laser beam obliquely incident on 
a graphene-n-InSb semiconductor rippled surface in the presence/absence of 
an external magnetic field. The possibility of generation of THz SMPs has been 
assessed within the
	frequency range of $\sim2-5$~THz, with the graphene Fermi energy 
spanning from 20 meV to 140 meV, and the external magnetic field from 0 
to 896.5 Gauss. Some important conclusions may be drawn from our results:

%\begin{enumerate}[(i).]
% \item aa
% \item bb
%
%\end{enumerate}
(i). The amplitude of normalized THz SMPs wave grows as graphene 
Fermi energy increases (Figure~\ref{Fig6}) because it is found from the 
dispersion curve of THz SMPs (Figure \ref{Fig3}) that
	the normalized propagation constant $K_z=k_{z}c/\omega_{p}$ increases with 
increase in graphene Fermi energy. 
When $k_{z}\gg\omega/c$, the terms in Eq.(\ref{eq14}) simplify as follows: $\alpha_{1} \approx k_{z}$, $\alpha_{2} \approx k_{z}$ and in the coupling term $-k_{z}^{2}$ dominates. Consequently, the amplitude of normalized THz SMPs can be expressed as
\begin{eqnarray}
%\nonumber
			\bigg| \frac{E_l }{E_0 } \bigg|  =  \bigg| \frac{ 
i \omega^2_{p} k_z^2 \times T_{tr} h d}{2 \omega
				\epsilon_{x x} {\left(  2 + \beta^2_1 
				 + \beta^2_2 \right)}} 
					\bigg[
			i ( {{\beta^{\ast}_1}  {+ \beta^{\ast}_2} 
			})\tilde{\text{v}}_{x} -2\tilde{\text{v}}_{z} \bigg]   \bigg|. \label{eq15}
		\end{eqnarray}
%\begin{eqnarray}
%\nonumber
%			\bigg| \frac{E_l }{E_0 } \bigg| & = & \bigg| \frac{ 
%i \omega^2_{p} k_z^2}{2 \omega
%				\epsilon_{x x} } {\left(  2 + \beta^2_1 
%				 + \beta^2_2 \right)^{-1}} 
%					\bigg[-2\tilde{\text{v}}_{z} 
%			+ i ( {{\beta^{\ast}_1}  {+ \beta^{\ast}_2} 
%			})\tilde{\text{v}}_{x}\bigg] \\ & \times & T_{tr} h d \bigg|. \label{eq15}
%		\end{eqnarray}
From Eq.(\ref{eq15}), it is evident that $|E_{l}/E_{0}|\propto k_{z}^{2}$. Therefore, the amplitude of the THz SMPs grows as the normalized propagation constant $K_{z}$ increases.

(ii). The field strength of THz SMPs wave grows as THz
	frequency $\omega$ increases (Figure~\ref{Fig5} and Figure~\ref{Fig6}) because 
required ripple wave-number $Q = q c/\omega_p$ increases (Figure~\ref{Fig4}) with 
increase in THz frequency $\Omega$. 

(iii).	The field amplitude of THz
	SMPs wave grows (Figure~\ref{Fig8}) as the temperature of 
the n-InSb semiconductor increases. This enhancement occurs because higher 
temperatures lead to an increase in the
	electron charge density $n_{0}$, which in turn increases $\omega_p \propto \sqrt{n_{0}}\propto \sqrt{T^{3/2}\exp(0.26/k_{B} T)}$~\cite{jing2022thermally,gao2023multifunctional}. As a result, the amplitude of THz SMPs wave grows which is consistent with Eq.(\ref{eq14}); i.e., $|E_{l}/E_{0}|\propto\omega_{p}^2\propto T^{3/2}\exp(0.26/k_{B} T)$.

(iv). The amplitude of generated THz SMPs wave reaches its 
maximum (Figure~\ref{Fig7}) at a normalized THz
	frequency where SMPs resonance occurs. 
	
	Our approach introduces a few extra tunable parameters such as graphene Fermi 
energy $\text{E}_\text{F}$, n-InSb semiconductor's
	temperature T, external magnetic field $B_0$ and incidence angle of laser $\theta$ for controlling the amplitude of THz SMPs wave. This scheme can be used
	to the development of actively tunable THz plasmonic devices, which hold significant potential for a wide spectrum of applications in communication
	technologies, sensing systems, and detection devices~\cite{liu2024high,akyildiz2022terahertz,10.1002/smll.202401151,liu2024terahertz}.

\vspace{-0.25cm}

	\section{Author Declarations}
\vspace{-0.25cm}

	\subsection{Conflict of interest} 
\vspace{-0.5cm}	
	The authors have no conflicts to disclose.
	
\vspace{-0.5cm}	

	\subsection{Author Contributions} 
	
\vspace{-0.25cm}

	Rohit Kumar Srivastav (RKS) and 
Mrityunjay Kundu (MK)
	have developed the work. 
	Figures and the original draft of the manuscript have been prepared by RKS. 
Review and editing of the final manuscript have been done by MK.

\vspace{-0.5cm}
	\section{Data Availability}
\vspace{-0.25cm}
	The data that support the findings of this study are available from the corresponding author upon reasonable request.

\vspace{-0.5cm}
\section{References}
\vspace{-0.25cm}
\nocite{*}
\bibliography{Biblography}% Produces the bibliography via BibTeX.
\end{document}